\documentclass[12pt]{iopart}
\usepackage{graphicx}

\begin{document}

\title[Thermodynamics of as-grown ASI]{On thermalization of  magnetic  nano-arrays at fabrication}

\author{Cristiano Nisoli}

\address{Theoretical Division, MS B258, 
\\ Los Alamos National Laboratory, Los Alamos, NM, 87545, USA}
\ead{cristiano@lanl.gov}
\begin{abstract}
We propose a model to predict and control the statistical ensemble of magnetic degrees of freedom in Artificial Spin Ice (ASI) during  thermalized adiabatic growth~\cite{Wang,Morgan}. We predict that  as-grown arrays are controlled by the temperature at fabrication  and by their lattice constant, and that they can be described by an effective temperature. If the geometry is conducive to a phase transition, then the lowest temperature phase is accessed in arrays of lattice constant smaller than a critical value, which depends on the temperature at deposition. Alternatively, for arrays of equal lattice constant, there is a  temperature threshold at deposition and the lowest temperature phase is accessed for fabrication temperatures {\it larger rather than smaller} than this  temperature threshold. Finally we show how to define and control the effective temperature of the as-grown array and how to measure critical exponents  directly. We discuss the role of kinetics at the critical point, and applications to experiments, in particular to as-grown thermalized square ASI, and to magnetic monopole crystallization in as-grown honeycomb ASI.
\end{abstract}

\maketitle

\section{Introduction:  artificial spin ice and its ground state(s).}
\subsection{Artificial spin ice}
The study  of frustration, degeneracy and entropy in  artificial materials which can be taylor-designed to desired specifications is a novel  trend both in magnetic~\cite{Wang, Tanaka, Cumings} and colloidal systems~\cite{Shokef2, Shokef}. Artificial Spin Ice (ASI)  is a two-dimensional array of elongated, magnetically interacting, single-domain, permalloy nano-islands whose shape anisotropy defines Ising-like spins arranged along the sides of a regular lattice~\cite{Wang}. Unlike naturally occurring magnetically frustrated spin ice pyrochlores ~\cite{Ramirez,Bramwell}, ASI allows direct imaging of its microstate, therefore providing a precise experimental benchmark to theoretical treatments. Since its introduction, ASI has been employed successfully to study frustration~\cite{Wang}, and extension of thermodynamics to granular systems governed by non-trivial interactions~\cite{Nisoli,Ke, Nisoli2}, topological defects~\cite{Morgan} and information encoding~\cite{Lammert}; it has also become a preferred ground for direct imaging of a new striking fractionalization phenomenon: Òmagnetic monopolesÓ~\cite{Castelnovo,Jaubert,Morris}.
\subsection{Magneto-fluidization and real thermalization}
The dimensions of the islands which compose ASI vary somewhat between the different physical realizations. The choice of Morgan {\it et al}~\cite{Morgan} is rather typical (and not much different from the ASI of Wang {\it et al} ~\cite{Wang}):  $280\times 85$ nm$^2$ for the surface, with a height  of $16$ nm, arranged on a square lattice of lattice constant $a=400$ nm, which returns magnetic interactions on the order of $10^4-10^5$ K. Clearly thermal fluctuations cannot induce ``spin'' flips, and the material is static at room temperature. Therefore, early proponents, including the author, approached ASI as a complex granular material which could be externally driven via magneto-fluidization~\cite{Ke, Nisoli2}. The application of a rotating and time decreasing  magnetic field to ASI proved successful in lowering its energy and in returning a controlled variety of statistical ensembles  whose detail can remarkably be  predicted  in terms of an effective temperature~\cite{Nisoli2}. Yet, for the square ASI the protocol failed to realize--or even approach--its non degenerate ground state.

In 2010, in a novel approach to reach ASI's lowest energy state, Morgan and collaborators successfully reached what seems to be ASI thermalization during fabrication~\cite{Morgan}. They grew  square ASI via permalloy evaporation as very thin films on a pre-patterned substrate of silicon, and observed, through magnetic force microscopy, the formation of large crystallites of ground state domains, separated by domain boundaries, and containing only sparse topological defects. Magnetic monopoles in square ASI are energy excitations on top of the ground state, hence the interest in an approach which can reach that ground state. Obviously, control over the microstate of the as-grown ASI would be highly desirable. In this article, we propose  ways to achieve that control during fabrication. 
\section{Adiabatic growth}
\subsection{Assumptions}
We  model the thermalization at growth for ASI  of different geometries under the following assumptions:
\begin{itemize}
\item{The growth is adiabatically slow: at each instant the array is in thermal equilibrium.}
\item{At each instant during deposition, the height  $h$ of each island is about the same across the array.}
\item{As ASI grows, it crosses an energy region in which the magnetic interactions are on the order of the thermal energy or smaller.}
\item{At each stage of growth, there  is a blocking temperature below which the system freezes on a time-scale  commensurate with the deposition rate.}
\end{itemize}
\subsection{Blocking temperature}
We approach the problem from the point of view of superparamagnetism~\cite{Neel} in which the nano-islands are treated as  single-domain magnets, and  can randomly flip the direction of their magnetization  at temperatures larger than a {\it blocking temperature}. Single domain nano-islands have a volume-dependent, and therefore a height-dependent, energy barrier for spin flipping. We can therefore introduce a height-dependent blocking temperature $T_b(h)$, where $h$ is the height of the islands, and assume that when $T>T_b$ no spin flip takes place (see later for a discussion on the kinetics at stopping). $T_b$ can be computed via micromagnetic simulations for islands of any particular shape and dimensions, but in general we postulate
\begin{equation}
T_b(h)= \tau_1 A h,
\label{Tf}
\end{equation}
where $h$ is the height of the island and $\tau_1$ (a temperature per unit volume) only slightly depends on the area $A$, as cooperative internal relaxations soften the magnetic reversal. Since $\tau_1 \propto  M^2$, $\tau_1$ has a slight dependence on temperature through the density of magnetization $M$ from magnon contribution, which for permalloy we can neglect (introduction of that dependence on the following is trivial). As thermalization takes place at small $h$, on the order of a few nanometers, we take $T_b(h)$ linear in $h$\footnote{Deviations from linearity would include a negative correction $\propto -h^2$ to account for increased  internal relaxation during spin moment reversal in a taller nano-island, and would not affect qualitatively our treatment.} . 

As deposition increases it reaches a blocking point, after which the blocking temperature $T_b(h)$ is larger than the temperature $T$  at which deposition is performed, or $T_b\ge T$. The system then freezes in a thermodynamic state dependent on the deposition temperature $T$. From~(\ref{Tf}) we find
\begin{equation}
h^*(T)= \frac{T\label{hc}}{\tau_1A},
\label{hstar}
\end{equation}
for the blocking heigh, or the (average) eight of the islands when dynamics stops, for  deposition performed at temperature $T$. Obviously $h^*$ increases with $T$: larger temperature at deposition extends the dynamical range of ASI, delaying its freeze to higher depositions.  (In the following we will denote the value of observables at the blocking point with a star.)
\subsection{Effective temperature}
We have assumed the magnetic degrees of freedom of the array to be in thermal equilibrium when the blocking temperature $T_b$ crosses the deposition temperature $T$ and dynamics stops. Since the array freezes into a definite thermodynamic state t the blocking point, we can introduce an effective temperature $T^{\mathrm{eff}}$  as the temperature the observed ensemble would have in order to be Gibbsian in the as-grown energetics. 

In treatments of superparamagnetism,  interactions between magnetic nano-islands are often neglected~\cite{Knobel, Bedanta},  yet their role in determining the statistical ensemble of ASI is obviously fundamental. Let  $E(h)$ be any relevant energy emerging form inter-island interactions in an array whose islands have height $h$. Given the dipolar nature of the interaction, we assume that $E(h)$  scales as
\begin{equation}
E(h)= \epsilon (l/a) \frac{h^2}{a^3} A^2
\label{Eh}
\end{equation}
where $a$ is the lattice constant, $l$ is the length of an island, and we have assumed that $l \gg \sqrt A$, which corresponds to strong  anisotropy on the nano-island. Clearly $\epsilon(l/a)$ tends to a constant in the limit of $l/a\rightarrow 0$, the ideal dipole approximation, and is in general proportional to the square of the density of magnetization. 

If  $H$ is the final height of the islands when deposition is completed and $h^*$ the blocking height, then our assumptions allow us to define $T^{\mathrm{eff}}$  through the equation
\begin{equation}
E(H)/T^{\mathrm{eff}}=E(h^*)/T,
\end{equation}
which holds for any thermodynamically relevant energy $E$: indeed if the  system is  in equilibrium at the stopping point,  its thermodynamic ensemble is completely controlled by quantities like $E(h^*)/T$, and does not change after $h$ exceeds the blocking height  $h^*$, and while it grows from $h^*$ to $H$.
From (\ref{Eh}), the ratio between the energies is simply $E(h^*)/E(H)={h^*}^2/H^2$ and therefore one has that $T^{\mathrm{eff}}=T H^2/{h^*(T)}^2$. From (\ref{hstar}) we then  find our first result
\begin{equation}
T^{\mathrm{eff}}=\frac{\tau_1^2A^2 H^2}{T}.
\label{Teff}
\end{equation}
Equation (\ref{Teff}) shows an interesting fact: the effective temperature is  {\it lowered} when the deposition temperature is raised.
This result is only apparently counterintuitive, since larger temperatures during fabrication extend the dynamic region of ASI during growth, as already noted. 

The effective temperature in (\ref{Teff}) does not depend upon the lattice constant but only on properties of the single islands, as area, height and density of magnetization via  $\tau_1$ (as also in the  case of the effective thermodynamics for the AC demagnetization~\cite{Nisoli, Nisoli2}, where an effective temperature  describes instead a stochastic process out of equilibrium). 
{\it This does not imply that arrays of different lattice constants fabricated at the same temperature would belong to the same statistical ensemble}: the system is controlled by ratios of energy over temperature of the kind $E(H)/T\sim a^{-3}$, which increase at smaller lattice constant. Therefore arrays of smaller lattice constant $a$  belong to thermodynamic ensembles closer to the ground state. A less natural yet more general definition of effective temperature which might be more useful in experiments performed by varying the lattice constant will be given below.

%
\section{Crystallization}
\subsection{Critical lattice constant}
ASI of certain geometries are expected to undergo interesting phase transitions. In particular the square lattice should crystallize into an antiferromagnetic tiling. 
\begin{figure}[t!]   
\begin{center}
\includegraphics[width=3. in]{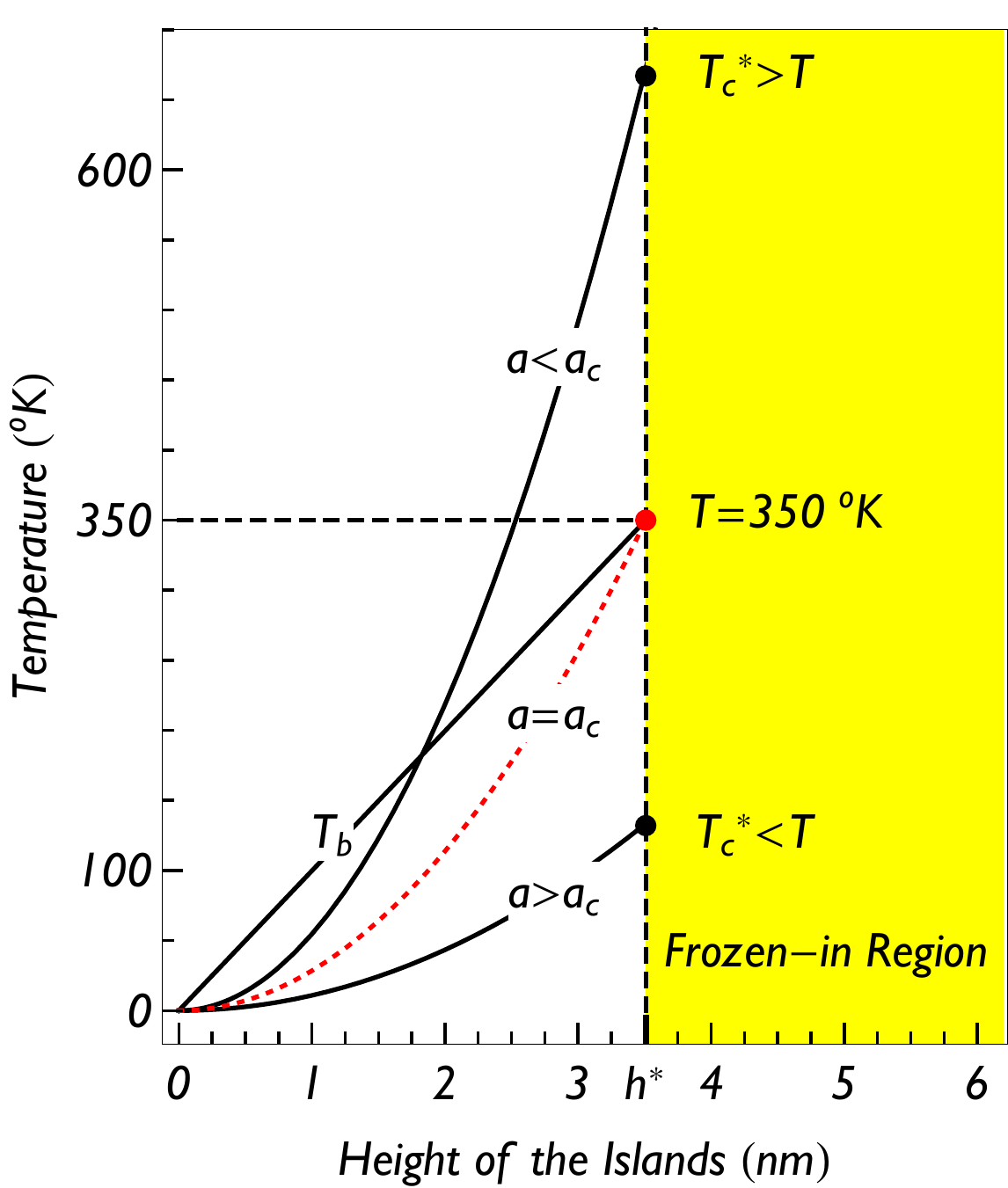}\includegraphics[width=3. in]{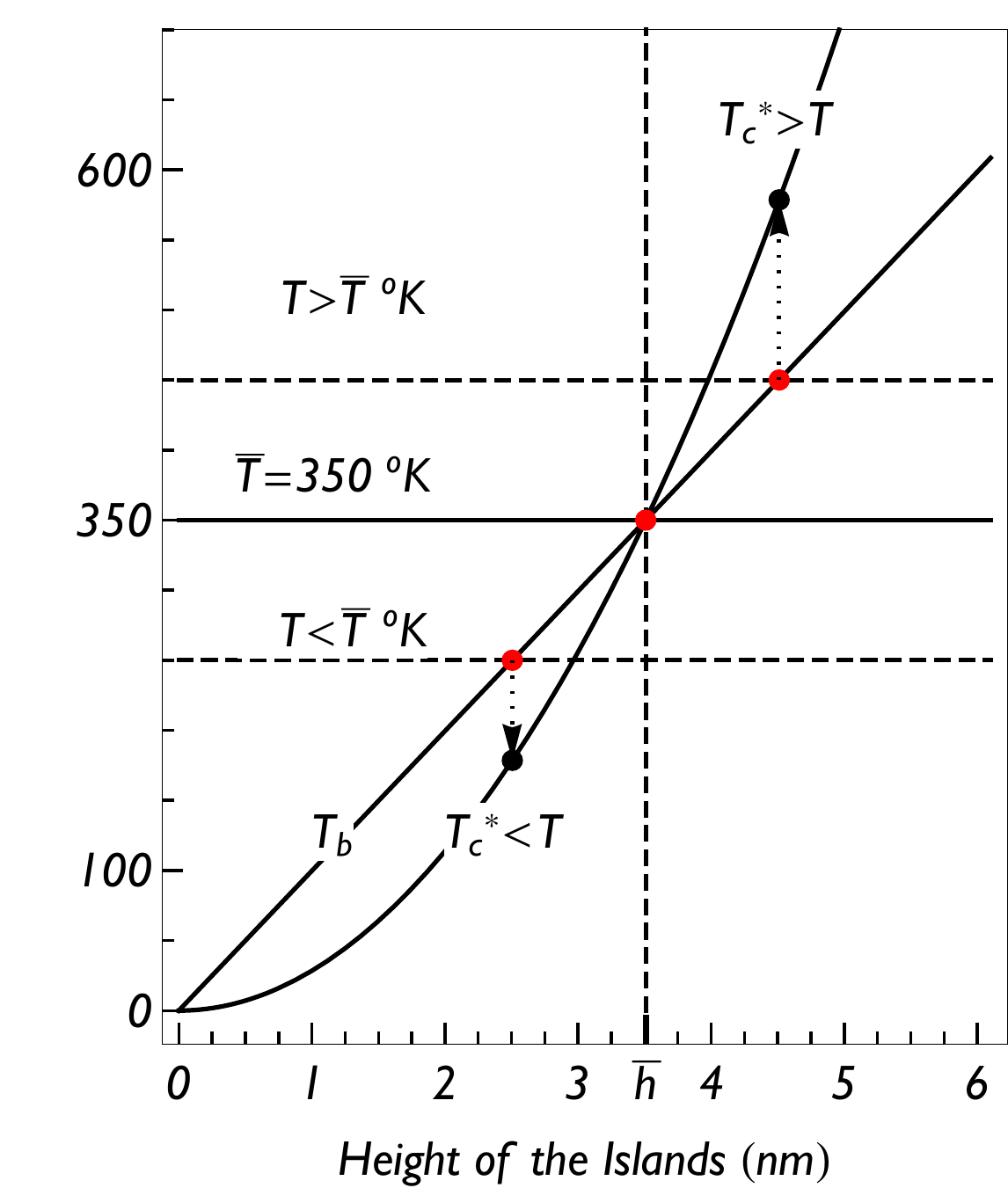}
\caption{Left: Arrays of different lattice constant grown at the same fabrication temperature show an ordered microstate when the lattice constant is smaller than a critical value $a_c$. In figure we plot the blocking temperature $T_b(h)$ , and the critical temperatures $T_c(h)$ for arrays of three different values of the lattice constant [$a>a_c$, $a<a_c$, $a=a_c$ (this last one in red)] as a function of the height of the nanoislands $h$. $h^*$ is the stopping height,  after which $T_b>T$ and therefore  freezing-in  starts. When $a>a_c$, $T^*_c<T_b=T$ at the blocking point, and therefore no crystallization is attained. When $a<a_c$, $T^*_c>T_b=T$ at the blocking point, and therefore crystallization is attained. $a_c$ is determined by the intersection $T=T_b=T_c$. For definiteness we have chosen $T=350$ K, $\tau_1=6.5~10^{-3}$ K nm$^{-3}$, $\tau_2=10$ K nm$^{-3}$, which return $h^*=3.5$ nm and $a_c=435$ nm.

Right: During fabrication of arrays of equal lattice constant the microstate can be controlled by varying the temperature $T$ at deposition. The figure shows the inverted temperature behavior, in which crystallization happens for deposition temperatures higher than the  temperature threshold $\bar T$, defined by intersection between blocking temperature $T_b[h]$  and critical temperature $T_c[h]$, or $T_b(\bar h)=T_c(\bar h)=\bar T$. When $T>\bar T$ the critical temperature for condensation (black dots) is larger than the deposition temperature (red dots) when the spin freeze-in, at the intersection of $T_c$ and $T$, and therefore the as-grown array  shows a crystallized microstate. The opposite happens for $T>\bar T$. For $T\sim \bar T$ then $T_c~T_b$ and kinetics effects must be taken into account. For definiteness we have chosen $\bar T$ to coincide with room temperature, although in Ref~\cite{Morgan} that is clearly not the case. }
\end{center}
\label{Fig1}
\end{figure}
Let $T_c$ be any critical temperature for a specific ASI geometry. It could be the  temperature for crystallization of square ASI  into its antiferromagnetic ground state. Or it could be the critical temperature of the ``Ice II'' phase predicted  for honeycomb ASI via numerical works~\cite{Moller, Oleg} and  correspond to the crystallization of magnetic monopoles of opposite charge on neighboring vertices, much like a NaCl ionic crystal. (For specificity, from now on we will talk of ``crystallization'' in a general sense whenever we allude to ASI undergoing any phase transition.) 

  $T_c$ is a particular case of (\ref{Eh}) and for an array of lattice constant $a$, comprising islands of surface $A$ and height $h$, it reads
\begin{equation}
T_c(h)=\tau_2\frac{h^2}{a^3} A^2.
\label{Tc}
\end{equation}
The same considerations exposed above for $\epsilon(l/a)$ apply now to $\tau_2(l/a)$, which like $\tau_1$ has the dimension of a temperature per unit volume and is proportional to the square of the density of magnetization of the material, $\tau_2\propto M^2$. 

Like $T_b$, $T_c$ also depends on the height $h$ of the islands, although quadratically rather then linearly. Figure~1.a plots $T_c$ and $T_b$ vs. $h$ for different lattice constants $a$. One can see that for large $a$, $T^*_c=T_c(h^*)<T$: therefore when the islands stop flipping at $h^*$,  the critical temperature $T^*_c=T_c(h^*)$ is lower than the experimental temperature $T$, and no crystallization has yet occurred. Instead the opposite happens for very small lattice constants. There exists therefore a temperature-dependent critical lattice constant $a_c(T)$ such that for lattice constants $a<a_c$  one expects crystallization, whereas for $a>a_c$ none is expected. 
From Figure~1.a the critical lattice constant $a_c$ is found by equating
$T^*_b=T^*_c=T$. Via (\ref{Tf}--\ref{Tc}) one finds 
\begin{equation}
a_c(T)=\sqrt[3]{\frac{\tau_2 T}{\tau_1^2}}.
\label{ac}
\end{equation}
Clearly $a_c$  depends on the deposition temperature  $T$ and larger temperature during deposition allows for larger arrays to reach crystallization. In the case of square ASI, arrays of lattice constant lower than $a_c$ are grown in a crystallized microstate, whereas arrays of larger lattice constant should show thermal disorder, while still being described by a Gibbsonian distribution, in terms of an effective temperature.

\subsection{Inverted temperature behavior}
Alternatively, arrays of identical lattice constant can be grown at different deposition temperatures. Then Figure~1.b  shows that if we call $\bar T$ the temperature at which the curves of $T_c(h)$ and $T_b(h)$ intersect, or $\bar T=T_c(\bar h)=T_b(\bar h)$,  growth of crystallized arrays correspond to a deposition temperature $T>\bar T$.  Indeed, when the deposition temperature is higher, or $T>\bar T$,  the critical temperature at the blocking point is larger than the temperature at which the experiment is performed, or $T^*>T$, and the system has already undergone crystallization when dynamics freeze. Conversely, growth at temperature $T<\bar T$ results in a thermally disordered array at the moment in which dynamics stop. Effectively, $\bar T$ represents a critical temperature, corresponding to $T_c(\bar h)$ but, interestingly, the system crystallizes above rather than below that critical temperature. To avoid confusions with the height dependent critical temperature of the arrays, we call $\bar T$  the {\it temperature threshold} at fabrication.
\begin{figure}[t!]   
\begin{center}
\includegraphics[width=2.8 in]{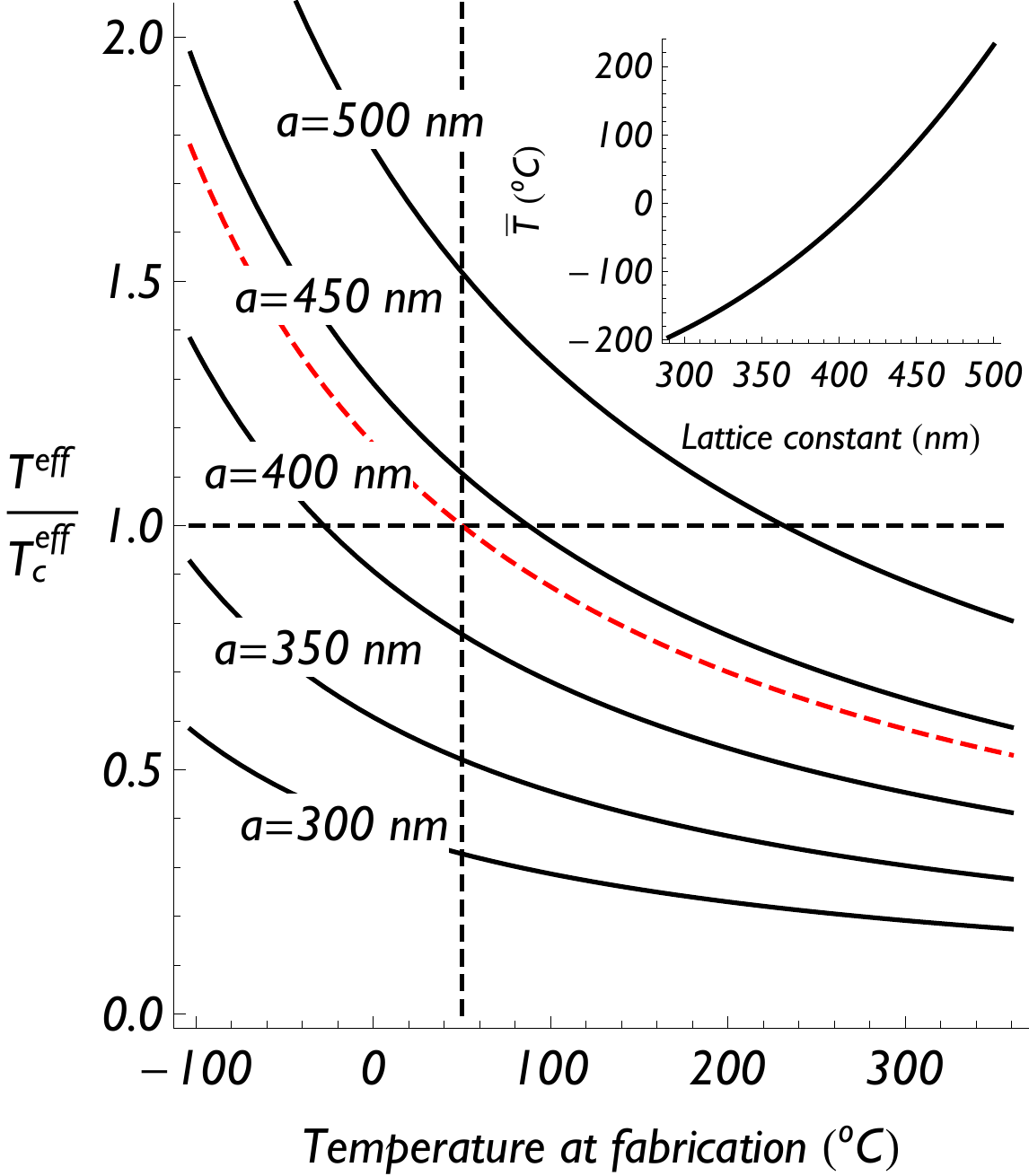}\hspace{5 mm}\includegraphics[width=2.8 in]{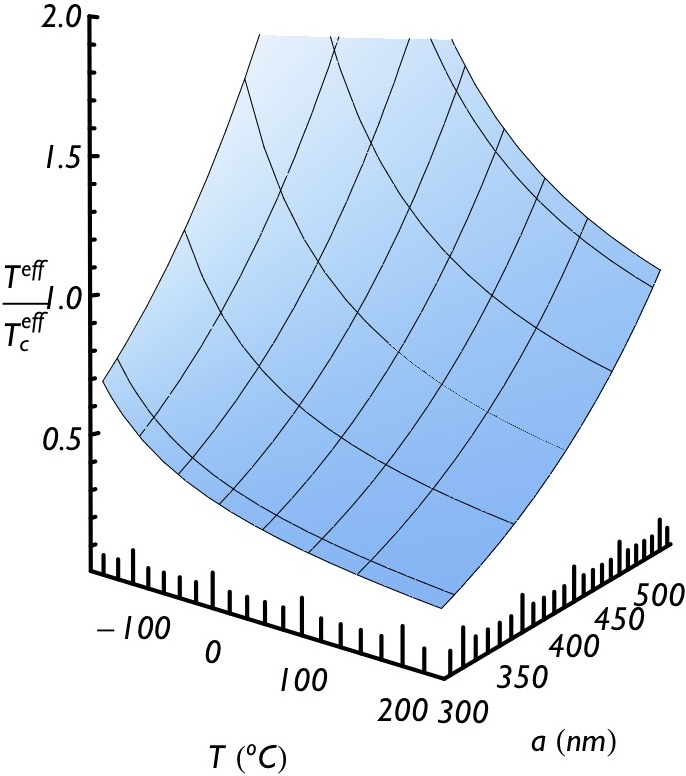}
\vspace{5 mm}
\caption{Left: Plot of $T^{\mathrm{eff}}/T^{\mathrm{eff}}_c$, the effective temperature of an as-grown array, measured in units of the critical effective temperature as a function of fabrication temperature $T$ in Celsius, for different values of the lattice constant $a$, from 500 nm to 300 nm,   from (\ref{trel}). The dashed vertical line represents the temperature of ~50 Celsius estimated in the experiments of Ref~\cite{Morgan}.  For definiteness we have chosen $\tau_1=6.5~10^{-3}$ K nm$^{-3}$, $\tau_2=10$ K nm$^{-3}$, which at $T=350$ K return the blocking height $h^*=3.5$ nm and the critical lattice constant $a_c=435$ nm (dashed red line), likely close to the experimental case of Ref~\cite{Morgan}. Inset: the  temperature threshold $\bar T$ for crystallization for arrays of different lattice constant. Right: For perspective,  curves described on the left are drawn as cuts on  the surface plot of  $T^{\mathrm{eff}}/T^{\mathrm{eff}}_c$ vs. the fabrication temperature $T$ (Celsius) and the lattice constant $a$ of the array (nm). } 
\end{center}
\label{Fig2}
\end{figure}

This inverted temperature behavior is  a consequence of the larger dynamical regime afforded by larger fabrication temperature. 
 $\bar T$ can be computed easily as
\begin{equation}
\bar T=\frac{\tau_1^2  a^3}{\tau_2},
\label{bT}
\end{equation}
and is a function of the specific material and the geometry, and proportional to the square of the density of magnetization, $\bar T\propto M^2$. If $T^*_c(T)$ is the value of the critical temperature for the array at the blocking point for different values of the fabrication temperature $T$, from (\ref{hc}, \ref{Tc}, \ref{bT}) we have 
\begin{equation}
\frac{T}{T^*_c(T)}=\frac{\bar T}{T},
\label{trel}
\end{equation}
which shows that {\it theoretically} the ground state can only be achieved  for infinite  temperature at deposition. In practice, crystallization  detectable by the limited size of MFM images can be typically achieved for reasonable values of  $(T_c-T)/T$, typically larger than  $0.1$. From (\ref{trel})
\begin{equation}
\frac{T^*_c(T)-T}{T}=\frac{T-\bar T}{\bar T}.
\label{Delta}
\end{equation}
Equations (\ref{trel},\ref{Delta})  show again that larger temperature at fabrication  leads to a more ordered condensate. Equation (\ref{Delta}) is useful in relating a quantity fundamental for diverging observables at the phase transition to the temperature at fabrication, therefore allowing for direct extraction of critical indices, using the technique introduced by Lammert and collaborators to  directly  extract entropy from ASI~\cite{Lammert}.
\subsection{Critical point and slow dynamics}
We have shown that different phases  can be obtained for fabrication  temperature $T$ larger or smaller than the temperature threshold   $\bar T$ in (\ref{bT}). But what happens when $T\sim \bar T$ and therefore (FIgure 1.b) the blocking temperature is close to the critical temperature, or $T_b\sim T^*_c$?  The dynamical response of the array slows down right when  the system undergoes transition, and phenomena similar to those due to rapid cooling might be expected. Clearly it is now time for a few It is now time for a few kinetic considerations. 

There are three characteristic rates in our problem. One is the rate of thermally induced magnetization reversal for each island, which in the limit of non-interacting islands  is given by the N\'eel-Arrhenius law, 
$
\nu=\nu_{0} \exp\left(-E_b/kT\right)
$
where $E_b$ is the energy barrier for magnetization reversal, and $\nu_{0}$ is the Arrhenius pre-factor (typically $\nu_{0}\sim10^{10-12}$ $s^{-1}$ \cite{Brown}).  Then there is the relaxation time for lattice equilibration, whose rate we call $\nu_e$. Equilibrium is regained by flipping a certain number of spins (per unit area) and it is then reasonable to take $\nu_e\propto \nu$, or
\begin{equation}
\nu_e= \nu_{e,0}\exp\left[-E_b(h)/kT\right].
\label{nue}
\end{equation}
Finally there is the deposition rate $\nu_h$, given by the number of layers deposited in the unit of time. 

For each subsequent layer, the change in the energetics of the system corresponds to a small deviation from equilibrium.  But as deposition increases, so does the time $\nu_e^{-1}$ needed for the system to respond and re-equilibrate into a new state. Eventually the array response  is too slow to catch up with the change in energetics, and the process converges to a state close to the equilibrium ensemble. If that state is far from a phase transition, we can follow the general approach in superparamagnetism and approximate it to an equilibrium state whose temperature is given by  $\nu_{e}(T_b)= \nu_h$. With that choice, $T_b$ in (\ref{Tf}) is related to the coercive energy barrier $E_b$ via
\begin{equation}  
T_b(h)=\frac{E_b(h)}{\ln \left(\frac{ \nu_{e, 0}}{\nu_h}\right) }.
\label{Tf2}
\end{equation}
$E_b$ is   generally taken to be independent of the temperature $T$ and proportional to the volume. Equation (\ref{Tf}) is therefore justified with
$
\tau_1\propto{M^2}/{\ln \left(\frac{ \nu_{e, 0}}{\nu_h}\right) }
$, 
where $M$ is the density of magnetization of the material. The typical deposition rate in experiments is $\nu_h\sim 10^{-1}$~s$^{-1}$ much smaller than $\nu_{e, 0}\propto\nu_{0}\sim 10^{10-12}$~s$^{-1}$. Therefore, from an experimental perspective, changing the deposition rate even by a few orders of magnitude has negligible effect on $\tau_1$, and therefore on our predictions above.

This description  can break down when the blocking point is close to a critical point, since the change in the microstate during relaxation can in principle be dramatic. $h_c$ is the height at which the system undergoes the phase transition (and is therefore defined implicitly as $T_c(h_c)=T$). When it is much smaller than the blocking height $ h^*$ the array undergoes the phase transition before freezing. But when $T\sim \bar T$ then $h^* \sim h_c \sim \bar h $.  There is a narrow window around the stopping height $h^*$ in which the relaxation time of the system is one order of magnitude or less smaller than its value at $h^*$. With our choice of $T_b$ in (\ref{Tf2}), and with $h^*$ from (\ref{hstar}), we obtain  from (\ref{nue})  the size of that window 
\begin{equation}
\frac{h^*-h}{h^*}<\frac{1}{\log_{10}\left(\nu_{e,0}/\nu_h\right)}\simeq 0.1.
\label{glass}
\end{equation}
When $h_c$ lies below that window, the system undergoes the phase transition.
Naturally, the relative values of $h_c$ and $h^*$ cannot be chosen freely, but depend on the temperature at fabrication. From (\ref{Tf}, \ref{Tc}, \ref{bT}) we obtain
\begin{equation}
\frac{h^*-h}{h^*}=\frac{T-\bar T}{2T}.
\end{equation}
Therefore no kinetic concern should involve the phase transition when
\begin{equation}
\frac{T}{\bar T}=\frac{\tau_2 T}{\tau_1^2  a^3}>\theta=\frac{\log_{10}\left(\nu_{e,0}/\nu_h\right)}{\log_{10}\left(10^{-2}\nu_{e,0}/\nu_h\right)}\simeq1.2.
\label{glass2}
\end{equation}
When instead $T<\theta \bar T$, the system responds slowly at the deposition threshold for crystallization $h_c$, which might correspond to super-cooling or to glassy behavior. Notice that (\ref{trel}, \ref{glass2}) imply that when kinetic effects are negligible we have $T/T^*_c(T)<\theta^{-1}$ and therefore no glassy behavior or lack of equilibration can be induced far below the critical point. 

\section{Implications for as-grown ASI}
\subsection{Effective critical temperature}
The presence of an equilibrated ground state can be revealed by directly imaging the microstate of ASI, and our framework can be tested by mapping different microstates from arrays obtained at different deposition temperatures or of different geometry. 
It is therefore useful to relate the above considerations to measurable quantities in the as-grown samples. 

As mentioned above, the as-grown array corresponds to a specific frozen-in thermodynamic state, and therefore it is useful to express our results in terms of the effective temperature. From (\ref{Tc}) it follows that the critical temperature for an as-grown thermalized array with islands of height $H$ is 
\begin{equation}
T^{\mathrm{eff}}_c=\tau_2\frac{H^2}{a^3} A^2,
\label{TcH}
\end{equation}
 which we call effective critical temperature. By direct substitution, it is easy to prove equations analogous to (\ref{trel}, \ref{Delta}) in terms of the effective temperature:
\begin{equation}
\frac{T^{\mathrm{eff}}}{T^{\mathrm{eff}}_c}=\frac{\bar T}{T}=\frac{\tau_1^2  a^3}{\tau_2 T}
\label{trele}
\end{equation}
and 
\begin{equation}
\frac{T^{\mathrm{eff}}_c(T)-T^{\mathrm{eff}}}{T^{\mathrm{eff}}}=\frac{T-\bar T}{\bar T}.
\label{Deltae}
\end{equation}
Again (\ref{trele}) shows how the effective temperature can be reduced by increasing $T$ at fabrication, or by using lattices of smaller lattice constant. Figure 2 shows the behavior of ${T^{\mathrm{eff}}}/{T^{\mathrm{eff}}_c}$ for different fabrication temperatures and lattice constants under reasonable assumptions for $\tau_1$, $\tau_2$. For definiteness we have chosen $\tau_1=6.5~10^{-3}$ K nm$^{-3}$, $\tau_2=10$ K nm$^{-3}$, to return $h^*=3.5$ nm and $a_c=435$ nm at a deposition temperature  $T=350$~K. 

\subsection{Most general effective temperature}

The previous definition of effective temperature, as the temperature which the as-grown system would have to belong to the observed observed experimental ensemble at its as-grown energetics, is the most natural. It is also well suited for experiments in which temperature at fabrication can be controlled. From an experimental point of view, it is interesting to introduce a more general effective temperature which can take into account more directly of changes in lattice constant. 

In general, as explained above, the thermodynamic state is controlled by the quantities $E(h^*(T))/T$, where $E(h)$ is any relevant energy for the system of height  $h$ and is given by (\ref{Eh}). Therefore, from (\ref{hstar}, \ref{Eh}), and taking that $\tau_1\propto M^2$, $\epsilon\propto M^2$, where $M$ is the density of magnetization, we can introduce that the most general choice of an effective temperature $\tilde T^{\mathrm{eff}}$, as the one normalized to an energy scale independent from the variables of the problem, or

\begin{equation}
\tilde T^{\mathrm{eff}}\propto \frac{M^2  a^3}{T}.
\label{Te2}
\end{equation}
Clearly (\ref{trele}, \ref{Deltae}) hold for this alternative definition of effective temperature as well. 
\subsection{Disordered as-grown square ASI}

Let  $n_{\alpha}$ be the relative occurrence of a state of the system. Using (\ref{Te2}) one can compute
 $n_{\alpha}$ as function of different temperatures at fabrication, different lattice constants and for materials of different magnetization, via the Gibbs distribution
 \begin{equation}
 n_{\alpha}=Z^{-1} \exp\left(\chi_{\alpha} T/ M^2 a^3\right), 
 \label{na}
 \end{equation}
 where $\chi_{\alpha}$ is a constant and depends only of the scale invariant geometry, and the particular state $\alpha$.   
 
Direct visualization of as-grown square ASI obtained via slow and careful deposition of permalloy (see introduction)~\cite{Morgan} has revealed formation of crystallites of long range order. 
From (\ref{ac}) one expects  that a thermally disordered picture can be regained for larger lattice constants (lower temperatures would be impracticable, since $h^*$ for those experiments seems already rather low, on the order of a few nanometers).  Since thermal disorder might warrant an approach in terms of a gas of independent vertices~\cite{Nisoli, Nisoli2}, $ n_{\alpha}$ in  (\ref{na}), where $\alpha$ labels different vertex configurations~\cite{Wang},  can be used to predict the relative abundance of different vertices in depositions performed at different  lattice constant or temperatures.

\subsection{Crystallites in as-grown Square ASI}

The presence of grain boundaries in orderd as-grown arrays has been ascribed to inherent disorder in inter-island interactions~\cite{Kohli}, a phenomenon also seen in numerical simulations of  different systems at zero temperature~\cite{Charles1,Charles2}. 
Recent  numerical and experimental work on magneto-fludisation of square ASI has shown that disorder in the nonuniform energy barriers for magnetization reversal  leads to nucleation sites for ground state  crystallites of opposite orientation, rendering a single domain ground state unattainable~\cite{Budrikis}. In the case of as-grown thermalized ASI, our assumption of equal eight of the islands at growth, although  plausible, might neglect a disordered distribution of heights which in turn could provide similar nonuniform energy barriers for magnetization reversal. 

Even though the disorder-based mechanism for grain boundary formation  is not well understood in this case, it is reasonable to assume that if indeed it is the disorder in inter-island interactions to be responsible of the observed fragmentation, then a smaller lattice constant would lead to larger crystallites. Indeed the size of the crystallites at zero temperature is likely determined by the ratio between the energy cost of the grain boundaries and the energy variations in inter-island interactions due to quenched disorder in the size and shape of the nano-islands. As lower lattice constants increase the inter-island interaction without changing the energy disorder, it might lead to larger crystallites. A proper thermodynamics of quenched disorder in square ASI and its effect on crystallite formation could be employed within our framework to predict crystallite size in as-grown ASI, via the effective temperature in  (\ref{Te2}). If instead fragmentation is mostly consequence of the disorder in the coercive barrier of the islands due to non uniform height $h$ during deposition, which  reflects in non-uniform blocking temperatures, then our treatment would suggest that larger temperatures at deposition should return crystallites of larger size. In fact a larger fabrication temperature $T$ leads to larger stopping heigh $h^*$ and therefore to a reduced relative magnitude of  disorder in $\Delta h/h^*$ at the blocking point.

Our approach suggests that In addiction to disorder one might consider another source of fragmentation  into subdomains. As explained above, when $\bar T<T<\theta\bar T$, kinetic effects become important at crystallization. Real materials are known to crystallize into domains of different orientations when cooled at a fast rate, and the same phenomenon could be taking place in as-grown square ASI. We do not know the value of the temperature threshold $\bar T$
\footnote{From our purely illustrative  choice of $\tau_1=6.5~10^{-3}$ K nm$^{-3}$, $\tau_2=10$ K nm$^{-3}$, used in the Figures, and $a=400$ nm, we get, from (\ref{bT}),  $\bar T=270$ K.  Since $T=350$ K, we have $T/ \bar T=1.3$, which, according to  (\ref{glass2}), lies just above the kinetic window. Of course,  a slightly different yet equally reasonable choice of those parameters  returns a ratio of$T/\bar T$ corresponding to the region of slow dynamics, suggesting that  kinetic effects might play a role.} 
in the experiments of ref~\cite{Morgan}, yet if the formation of crystallites is indeed a consequence of  the  proximity of the crystallization point to the blocking point then from (\ref{bT}, \ref{glass2}) an increase of say 30\% in the temperature at deposition, or a reduction of 10\% in the lattice constant, should take the critical point out of the kinetic window and therefore considerably change the size of crystallites. 

\subsection{Dynamical ASI}

It would be interesting to fabricate more dynamical ASI which could then be equilibrated at different temperatures. Arrays responsive to thermal fluctuations might be obtained by playing with the magnetization of the material but also via deposition techniques, by keeping $H$, the final height of the islands, small, yet not {\it  too} small.  Figure~1.b shows that when $H<\bar h$ the resulting as-grown ASI will never be able to approach the region of critical temperature which, depending on the application, might or might not be desirable: indeed when exposed to temperatures lower than $T_b(H)$, ASI would not respond, and since $T_c(H)<T_b(H)$ if $H<\bar h$ then the phase transition would be inaccessible to thermalization. $\bar h$ can be easily computed as
\begin{equation}
\bar h=\frac{\tau_1 a^3}{\tau_2 A}
\label{bh}
\end{equation}  
and interestingly does not depend on the density of magnetization.

For square ASI, ref~\cite{Morgan} shows that crystallization is achieved and since they report a blocking height of a few nanometers,  we can take $\bar h\sim 1$~nm for their system. Doubling the lattice constant would give, from (\ref{bh}),  $\bar h \sim 10$~nm. Then an array of $a=800$~nm and  thickness $H\sim 5$~nm would then respond to external temperature, even at room temperature, while  always being  disordered. However an array of  $a=400$~nm and thickness $H\sim 5$~nm  would also equilibrate and respond to external temperature, while also accessing an ordered phase. 

\section{As-grown ASI and magnetic monopoles}

Magnetic monopoles were introduced in naturally occurring spin ice pyroclore to subsume the effect of long-range interactions in a simple description of  low energy excitations~\cite{Castelnovo}. The role of their Coulomb-like charge is confirmed by the explanation of low-temperature behavior of spin ice in terms of a liquid-gas transition of monopoles. 

Magnetic monopoles have been directly observed in honeycomb ASI. Yet  while these topological excitations do correspond ``structurally'' to the magnetic monopoles of spin ice, insofar as one can somehow formally attribute a ``net magnetic charge'' to the excited vertices, the effects of their long range interaction is still unclear.  To rightfully deserve their name in ASI, monopoles must be shown to provide a similar low-energy description, amenable to thermodynamic treatment.  

Unlike magneto-fluidised ASI, which returns higher energy macro-states, thermalized as-grown ASI provides ordered states, in which magnetic monopoles could describe low energy excitations. By controlling temperature and lattice constant as explained in our approach, as-grown thermalized ASI can map microstate probabilities corresponding to different effective temperatures and provide a promising playground to test thermodynamic treatments of magnetic monopoles. This would finally  assert their {\it reality} as point-like, long-range interacting excitations. Below we propose directions to achieve this goal within our framework.

\subsection{Square ASI and monopole excitations}

Morgan {\it et al.}~\cite{Morgan} witnessed the formation of local excitations inside ordered crystallites, computed their energy numerically via a point dipole model, and showed that their relative frequencies follow a Gibbsian distribution, which further corroborates the idea that real thermalization is taking place during growth. They also pointed to particular defects in the form of monopole charges  connected by energetically costly Dirac strings (or more properly Nambu strings~\cite{Nambu,Mol0}) and noticed their tendency to form closed configurations with the string looping, rather than configurations with long open strings. They interpreted this as an effect of the monopole-antimonopole long range magnetic attraction. 

It would be interesting to raise the effective temperature (\ref{Te2}) by lowering the fabrication temperature, to see whether the change induces an opening of such loops and more separated monopoles, and if a description of their energies in terms of a Coulomb interaction can provide faithful predictions of their relative abundance via (\ref{na}).

M\'ol {\it et al.}, have predicted a monopole-unbinding transition ~\cite{Mol0,Mol, Mol2, Silva}, in which the entropy of the Dirac string overcomes its energy cost. This transition could also be investigated by fabricating ASI of different effective temperature. Although  numerical results predict the transition in a state of thermal disorder, which would prevent its direct observation in square ASI via magnetic force microscopy, that might be a consequence of the point dipole approximation employed by the authors. Indeed in a numerical work on honeycomb ASI M\"oller and Moesner have shown that, as one would expect, monopole signature becomes less observable when the ratio $l/a$ tends to zero~\cite{Moller}. Also, even if impossible to spot by eye in a MFM image, a transition could be seen by extracting the entropy for different effective temperatures with the method illustrated by Lammert~\cite{Lammert} and using it to compute the specific heat curve.

\subsection{Honeycomb ASI and crystallization of monopoles}

As mentioned above, our description of thermalization in as-grown ice is independent of the particular geometry and can be applied to honeycomb ASI as well. While as-grown square ASI could be the ideal candidate to study monopoles as sparse  excitations, honeycomb\footnote{A honeycomb ASI is made of nano-island arranged along the edged of a honeycomb pattern. It can be modeled  by dipoles on a kagome lattice or by monopoles on the vertices of an hexagonal  lattice. Since both theoretical description might apply to the same real material in different conditions, the author prefers to employ the more general nomenclature of ``honeycomb ASI'' rather than the more particular ``hexagonal ASI'' or ``kagome ASI'' to demarcate the actual physical material used in experiments from the different possible theoretical models that might apply to it.
} ASI is interesting in a different regard: it should reveal two phase transitions, recently explored numerically~\cite{Moller, Oleg}. At low temperature, a (pseudo) ice rule manifold appears, in which vertices manifest the 2-in/1-out or 2-out/1-in rule. Unlike the case of square ASI, in honeycomb ASI each low energy vertex is endowed by a positive or negative magnetic charge. Therefore a new phase transition (which  predicted numerically although not yet observed experimentally) should bring it to a lower energy configuration which Ref.~\cite{Moller} named Ice II, and which in practice corresponds to the crystallization of monopoles of opposite charges on nearest-neighbor vertices, therefore forming a triangular ionic crystal of monopoles. At even lower temperatures, an ordered phase emerges because of long-range dipolar  interactions neglected in the monopole approximation: the loop state. The two critical temperatures for these phases strongly depend on the ratio between the island length $l$ and the lattice constant $a$~\cite{Moller}, a property that can be exploited for intelligent fabrication.

Unlike the case of square ASI,  magneto-fluidisation~\cite{Ke,Nisoli2} successfully anneals the honeycomb ASI into its pseudo-ice manifold. Yet it fails to reveal any monopole crystallization. Magneto-annealed samples return an extracted entropy per spin of $\sim 0.75$~\cite{Lammert} rather than the $\sim .15$ expected at crystallization (taking the entropy of a random spin distribution to be 1), and extraction of nearest neighbors charge correlation from MFM immages provides values  $\sim.1$ rather than the crystallized value of $1$. 
It would therefore be very interesting to attempt as-grown thermalization of honeycomb ASI to investigate monopole crystallization and the loop state by controlling its effective temperature in the way described above and by extracting its entropy and computing monopole-monopole correlations. 
Our predictions above above apply, mutatis mutandis, to these two transitions as well, with a different choice of the constant $\tau_1, \tau_2$. Clearly in the case of monopole crystallization one expects a more complex dependence of those constants from the ratio $l/a$ as  shown in numerical calculations~\cite{Moller}. In particular, in the case $l\sim a$, because of the Coulomb interaction between magnetic monopoles, the dependence of  the critical temperature for monopole crystallization  will be $T_c(h)\sim h^2/a$ rather than the $\sim h^2/a^3$ of (\ref{Tc}).

\section{Conclusion}

We have treated the fabrication of ASI by slow deposition  as an adiabatic phenomenon and found that the probability of its microstate is described by an effective temperature which depends on the lattice constant of the arrays and the temperature at deposition. When a phase transition exists in ASI, then there is a geometry-dependent  temperature threshold such that the phase {\it below} the critical point can be achieved with fabrication temperatures {\it above} the temperature threshold. When the deposition temperature is close to the  temperature threshold, then kinetic effects due to the N\'eel-Arrhenius flipping dynamics are expected to play  a role similar to fast cooling at critical point. We have proposed how to employ these considerations to study monopole-unbinding in square ASI and monopoles crystallization in honeycomb ASI.

\section{Acknowledgments}

We thank Vincent Crespi, Paul Lammert (PSU), Jason Morgan  and Christopher Marrows (University of Leeds) for helpful discussions, and Tammie Nelson (LANL) for help with the manuscript.  This work was carried out under the auspices of {U.S.} Departmenf of Energy  at LANL under Contract No. DE-AC52-06NA253962, and specifically LDRD Grant 20120516ER.
 

\end{document}